# Urban Planning in the First Unfortified Spanish Colonial Town: The orientation of the historic churches of San Cristóbal de La Laguna

Alejandro Gangui and Juan Antonio Belmonte

**Alejandro Gangui**
Universidad de Buenos Aires, Facultad de Ciencias Exactas y Naturales, Argentina.
CONICET - Universidad de Buenos Aires, Instituto de Astronomía y Física del Espacio (IAFE), Argentina.

**Juan Antonio Belmonte**
Instituto de Astrofísica de Canarias.

**Abstract**
The city of San Cristóbal de La Laguna in the Canary Island of Tenerife (Spain) has an exceptional value due to the original conception of its plan. It is an urban system in a grid, outlined by straight streets that form squares, its layout being the first case of an unfortified colonial city with a regular plan in the overseas European expansion. It constitutes a historical example of the so-called "Town of Peace", the archetype of a city-republic in a new land that employed its own natural boundaries to delimit and defend itself. Founded in 1496, the historical centre of the old city was declared a World Heritage Site by UNESCO in 1999. We analyse the exact spatial orientation of twenty-one historic Christian churches currently existing in the old part of La Laguna which we take as a good indicator of the original layout of the urban lattice. We find a clear orientation pattern that, if correlated with the rising or setting Sun, singles out an absolute-value astronomical declination slightly below 20°, which, within the margin of error of our study, might be associated with the July 25[th] feast-day of San Cristóbal de Licia, the saint to whom the town was originally dedicated. We also discuss at some length some recent proposals which invoke somewhat far-fetched hypotheses for the planimetry of the old city and finish up with some comments on one of its outstanding features, namely its *Latin-cross* structure, which is apparent in the combined layout of some of its most emblematic churches.

## Introduction

Before the conquest, the area of the city San Cristóbal de La Laguna was a communal pasture region between the kingdoms (or *Menceyatos*) of Tegueste, Anaga, Güímar and Tacoronte and it was known by the aborigines as Aguere (Tejera 1991). It was a notable place to which they herded their cattle, seasonally, to take advantage of the fertile land. Aguere was the location of the most important battle that confronted the Castilian troops, under the command of Don Alonso Fernández de Lugo, and the aborigines under the command of Bencomo, *mencey* [king] of Taoro, in November 1494. The final victory in 1496 led to the annexation of the island of Tenerife to the Crown of Castile and historical sources locate that event close to July 25[th], feast day of San Cristóbal, which gave the city its founding name and patronage.

The location of the old town was chosen with strategic reasons in mind: its elevated terrain, away from the sea, connects the north and south slopes of the island, thus allowing its inhabitants control of the surrounding region and the protection of the city against eventual pirate attacks. It was also a unique plain, surrounded by mountains, with abundant water courses, and a small lagoon (in fact, Aguere was an aboriginal toponym meaning lagoon). It was, at the time, a very rich place, used by all the Menceyatos of the island as a communal area of grazing.

After the defeat of the aborigines, in the summer of 1496, the first feast of Corpus Christi was celebrated in the primitive and newly-built parish church of Our Lady of the Immaculate Conception [*Nuestra Señora de La Concepción*]. Located near the present building, the feast was an event considered to be the first foundation of the city. At that same time, the settlement of soldiers and civilians began in the old town, with the first modest huts being built at the foot of the church, at some distance from the lagoon that extended towards the north (and that would finally disappear in 1837, when it was drained). The initial nucleus was known as the Upper Town [*Villa de Arriba*], where an irregular layout of streets was developed as a result of a disorderly settlement, without prior distribution of the land (UNESCO 1999, 105).

But in the year 1500, Fernández de Lugo returned to the island after having capitulated to the Crown of Castile's demands regarding the appropriate conditions of the conquest, and carried out the definitive foundation of the city. He obtained from the Catholic Monarchs the title of Captain General [*Adelantado*] and the governorship of the territories. As such, he had the full right to administer justice, appoint the different administrative, judicial, and military positions, adjudicate land, dictate ordinances, and represent the head of the Town Council [*Cabildo*]. Let us recall that the first meeting of this institution was established on October $20^{th}$, 1497, and it was, for a long time, the principal and only governing body (Aznar Vallejo 2008, 191).

The definitive foundation of the city in 1500 was conceived from a new design centred around the current Adelantado's Square [*Plaza del Adelantado*], then known as the Lower Town [*Villa de Abajo*], about 1 kilometre to the southeast of the old town and away from the lagoon. From this new nucleus, a layout with an ordered grid arose which followed a distribution of straight streets based on the classic model. This was a pattern already used in some previously-founded cities such as those located in the lower Andalucía like Puerto Real and Santa Fe de Granada. But, in this case, it was a city without walls founded under new premises in conquered distant land, so it had to respond to new political, economic, and social needs. In the long run, this new city design would serve as a model for the colonizing process that would later take place in the new American territories (Navarro Segura 2006).

Thus, the city of San Cristóbal de La Laguna was, throughout the sixteenth century, configured into two population centres, which emerged, as we saw, in a different way. The first provisional settlement, chosen by the Adelantado, around the parish church of La Concepción, was characterised by not having a planned urban layout; just a few houses of rough stone, covered with straw that gave shape to a small hamlet. About three years later, around 1500, a second, more organised nucleus emerged, which promoted settlement of the population to the south and east of the territory. In fact, on April $24^{th}$ of that year, the Adelantado issued a decree by which the sale or construction of more houses in the old town was no longer allowed and determined that new constructions should be carried out "desde el l'espital de Santespiritus hazia el logar de Abaxo", that is to say, towards the southeast of the Holy Spirit or Convent of San Agustín (Alemán de Armas 1986, 15).

The historical centre of the city was practically defined towards the end of the sixteenth century, as is shown in one of the first preserved maps of the city (Figure 1), drawn by the Italian engineer, Leonardo Torriani in 1588, who had been sent by order of King Felipe II to study the military defense of the archipelago (Torriani 1978). Over the years, the two population centres, the upper and lower villas, combined into one.

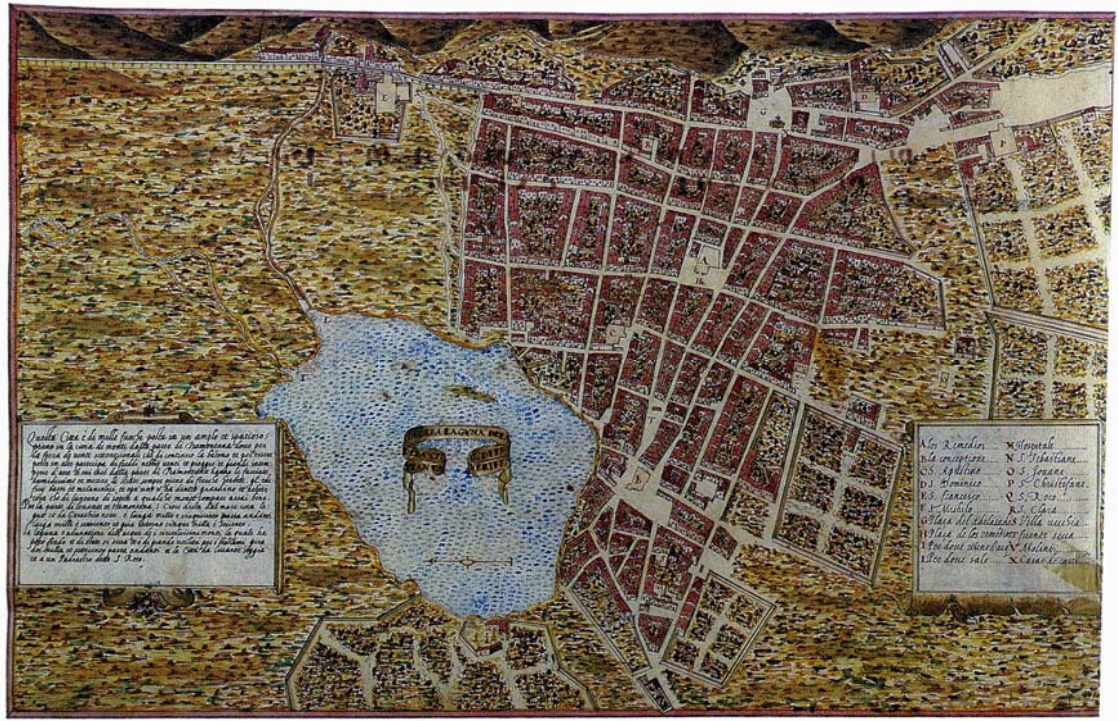

Figure 1: Map of the city of San Cristóbal de la Laguna towards the end of the sixteenth century, drawn by the Italian engineer, Leonardo Torriani. On this map, the north is approximately to the left. (This orientation of the map locates the Upper Town below the Lower Town.)

Although on the Old Continent, one can still find many cities with well-preserved historical centres which are much older than La Laguna, what makes this city of outstanding universal value is that it was not the result of centuries of interactions, modifications and reconstructions. Instead, San Cristóbal de La Laguna began to grow in a virgin, unpopulated, and remote place and, from the European perspective, was without history. What is unique about the city is that it is believed that it was conceived as a whole from the very beginning (the town as a project) and it is that precise original design, together with the morphology of the city and its remarkable urban profile, which has remained practically unchanged since the days of Torriani, more than 400 years ago.

From the very moment of the conquest of the Canary Islands by the crown of Castile, the arrival of the European population intensified, especially those coming from the coasts of western Andalucía. The migratory current that accompanied the first conquerors began its adaptation in the new lands and, undoubtedly, the origin of these people was decisive in the different architectural models that emerged in the newly-created settlements (Corbella Gualupe 2000).

This new architecture, in one way or another, was always tied to the economy of the Catholic Church. Where the ecclesiastical colleges had sufficient means, they opted for religious constructions endowed with arches and ogival forms. However, where the parishes were poor, or just beginning their collective life, as in the case of the Canaries, and, in particular, the Island of Tenerife, there was not much choice. This religious art had to be a popular art: simple, quick to create and economical (Figure 2). All these were characteristics that the Mudejar-style adequately met (Fraga González 1977).

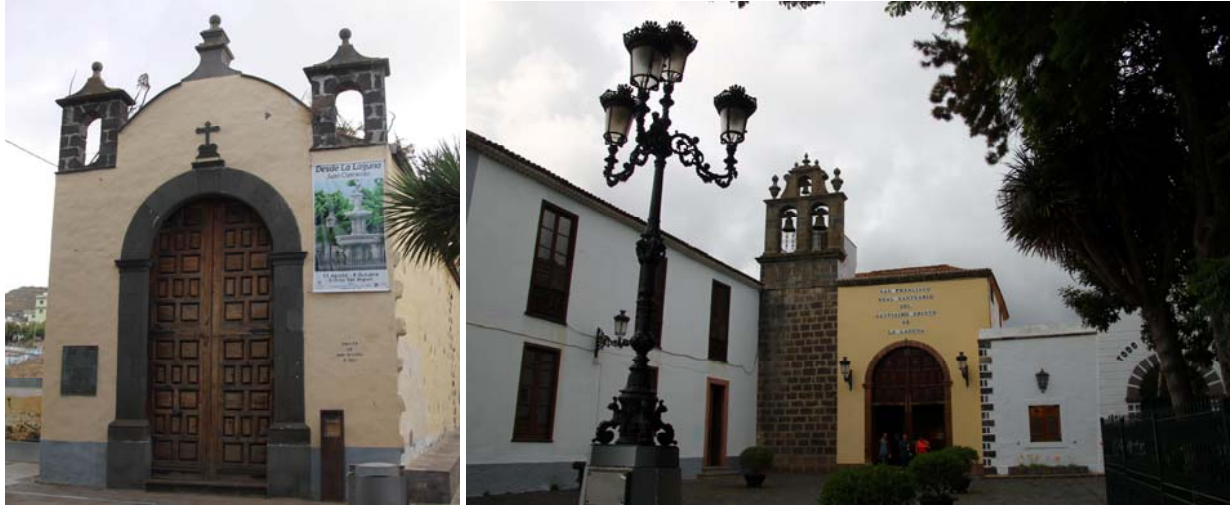

Figure 2: Two of the oldest constructions of the city. The San Miguel de los Ángeles chapel (left) was founded in 1506 by the Adelantado Fernández de Lugo, who expressed his wish to be buried there some day. The Royal Sanctuary of the Holy Christ of La Laguna (right), known as Saint Francis or El Cristo, was also founded early in the sixteenth century and suffered several major mishaps during its long history, such as floods and fires. Photographs by the authors.

The study of the orientation of old Christian churches has been of academic interest for a long time and has recently received new impetus in the literature, as it was recognised that orientation represents a key feature of the architecture of these churches. According to the texts of the early Christian writers and apologists, the symmetrical axes of religious buildings and churches' apses should lie along a particular direction, so that the priest had to stand facing eastward during services (McCluskey 2015, 1704).

North Africa, in spite of the Roman dominance, is an exception to this rule. In many regions, such as Proconsularis and Tripolitania, a number of old churches with orientations towards the west – which is a usual custom in the early times of Christianity – is found (Esteban et al. 2001, 81; Belmonte et al. 2007, 79). These regions are relevant to our study as they are possible homelands of the Canarian aboriginal population. Note also that most of these churches were oriented roughly within the solar range (with orientations between the winter and summer solstices) with noticeable clustering around the equinoxes and solstices (González-García 2015, 272).

In this work, we analyse the spatial orientation of most of the historic Christian churches currently existing in the old part of La Laguna. Our approach is based on the proposal that the alignment of churches can be considered a good indicator of the orientation of the original urban lattice. In our study, we also include a couple of religious constructions lying close to the borders of the city, as was depicted by Torriani in Figure 1, due to the fact that these churches date from previous times and, presumably, could be following the same orientation trend as the rest.

Our main aim is to find out whether a clear orientation pattern exists for the sample of churches (and therefore, according to our hypothesis, also for the layout of the city), and if this pattern can be correlated with the rising or setting Sun on particular dates throughout the year. As we will see, our results single out a broad interval, centred approximately on July 25$^{th}$ which, although nowadays corresponds to the feast-day of Saint James, in La Laguna, is still celebrated as the feast-day of San Cristóbal de Licia, the original patron saint of the city.

Before presenting our detailed analysis, in the next section, we will briefly discuss alternative ideas which assume a geometric origin for the planning of the city. Although we do not support this last view, essentially because it has no adequate historical support in documentary sources of the time, its introduction here is useful as it will serve to contextualise our project and to show that a simpler explanation for the particular layout of the city is worth pursuing.

**A city inspired by Greek philosophical principles?**

As we mentioned already, the exceptional value of the city relies in the original conception of its plan. It consists of a grid outlined by straight streets forming squares, giving, as a result, a layout which is the first-known documented case for an unfortified colonial city with a regular plan in the overseas European expansion. It constitutes a case study of the archetype of a city-territory that took advantage of its own natural boundaries to delimit and defend itself. In fact, its natural defensive systems have always been the Chamarta ravine to the southwest and the Gonzaliánes ravine towards the east border, at the foot of the slopes of San Roque, as well as the prominent lagoon towards the north, and other mountains by which it is surrounded (UNESCO 1999).

However, the city would be remarkable, not only for what has already been described in the previous paragraphs. According to popular modern tradition, La Laguna would be the material realisation of the first example of an unfortified colonial city, conceived and built according to "a precise geometric plan", with a solid basis on navigation and the science of the time, and whose symbolic structure could be interpreted in a manner similar to that of the marine charts or even the well-known constellations of the sky (Navarro Segura 1999, 121). Its map would have been organised according to a new peaceful social order, inspired by the religious doctrine of the millennium and expressed through the urban design, that arose in the year 1500 and would have been encouraged by the Catholic Monarchs themselves for the foundation of cities in the conquered lands (Navarro Segura 1999, 79).

The characteristics of the new foundation of the city can be attributed to the Adelantado himself, as a result of his stay in the Court, between August and October 1499, which at that time resided in the city of Granada. During those months, when Fernández de Lugo was preparing his future campaign in Barbary (North Africa), the Court supposedly dictated rules on how the process of the new foundation in La Laguna was to be carried out (Navarro Segura 1999, 204), according to the ideas rooted in humanism and religious renewal that prevailed at the time (the doctrine of the millennium already mentioned). From the comparative analysis between the future design of the actual city and the ideal city outlined in Plato's *Laws*, it is thought that the Catholic Monarchs agreed with the Adelantado "the application of the principles contained in the work [*Laws*] as part of an experimentation project of a new model of a city to be developed in the newly pacified territories incorporated to the Crown of Castile" (Navarro Segura 1999, 166).

On the basis of the bookish tradition in the Iberian Peninsula at the time, Navarro Segura speculates that Fernández de Lugo would have had Plato's book in his hands and, that from his careful reading of the text, he would have considered seriously the philosopher's ideas − and the eventual application of these ideas to La Laguna – with respect to the city of Magnesia discussed in Plato's dialogues (Navarro Segura 2006; Navarro Segura 1999, 186). In collaboration with Antonio de Torres, overseer of the Spanish monarchs for the African campaign, versed in mathematics and navigation, and with the Sevillian, Pedro de Vergara, mayor of the city of La Laguna for several years, the Adelantado would have carried out his design of the city during the first two decades of the new century. However, due to the premature death of Torres in a shipwreck in 1502 and the Adelantado's own disappearance in 1525, the ideas and symbolism behind this urban experiment would have been relegated to oblivion in the years that followed during the foundation of the American colonies (Navarro Segura 1999, 144).

Thus, in 1500, a new foundation of the city was in progress, which included the Upper Town and the new nucleus of population toward the southeast imposed by the Adelantado, that for many years remained physically separated. According to what has already been said, the new design for the city supposedly relied on ancient concepts based on mathematical formulas and delineated its streets through the use of nautical astrolabes and other navigational tools (Navarro Segura 1999, 121). From then on, Platonic and Renaissance concepts overlapped in the configuration of an original urban layout; a utopian and symbolic city, a projection of the sky on the earth that was oriented by the compass rose, and that had the convent of San Agustín (old limit of the upper villa, Figure 3) as its exact geometric centre.

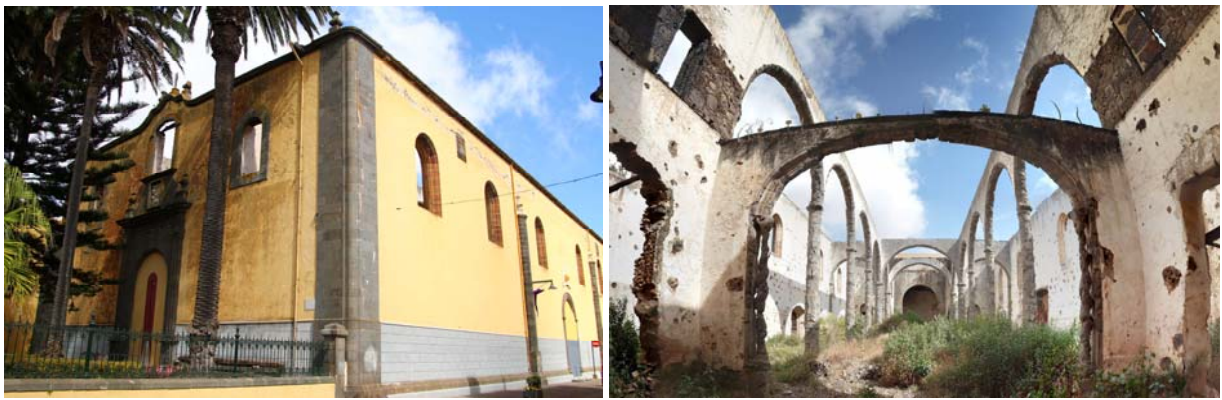

Figure 3: Main facade and walls of the church of San Agustín, next to the old convent of the same order, in its present state, after the fire of 1964 consumed its interior and collapsed the roof. Its location marked the south-eastern boundary of the old town, until the Lower Town was built. Photographs by the authors.

According to a study published by Navarro Segura, the design of the city was also based on the application of the instructions contained in Vitruvius's treatise *De Architectura* (c. 15 BC). These rules were simple to apply due to the common use of navigation-bearing guidance instruments and the detailed knowledge of the eight main winds of the time (Navarro Segura 1999, 204). In the pages dedicated to the foundation of cities, Vitruvius focuses on the winds as a key factor for the health of the different places and reduces to eight the set of regions on the horizon. The figure that best fits this design is the regular octagon inscribed in a circumference which would have been the one used in La Laguna where the Vitruvian octagon supposedly was employed to calculate the positions of the eight main winds and, therefore, to design the layout of the streets "midway", that is, along directions intermediate to those of these winds, which in the plain of Aguere were always particularly intense (Vitruvio 2010, 39).

This intentionality in the geometric layout of La Laguna would apparently be reinforced by some axes that cross the city, establishing particular distance relationships between representative religious buildings of the sixteenth century. It is pointed out that the geometric centre of San Agustín is equidistant from four old temples, taken in pairs: the church of San Juan Bautista and the church of El Cristo de La Laguna, on the one hand, and the chapels of San Roque and San Cristóbal, on the other (Navarro Segura 1999, 223). In addition, the religious axis linking the church of San Benito Abad (also of the sixteenth century) with the centre of the map in San Agustín, would be arranged precisely in a west-east direction, and divides the city-octagon into two parts. Finally, the distance separating the centre from the first two churches would be the same as that separating it from the two ends of the street that was once called Water Street [*Calle del Agua*], now Nava y Grimón back street and eastern boundary of the city (Figure 4).

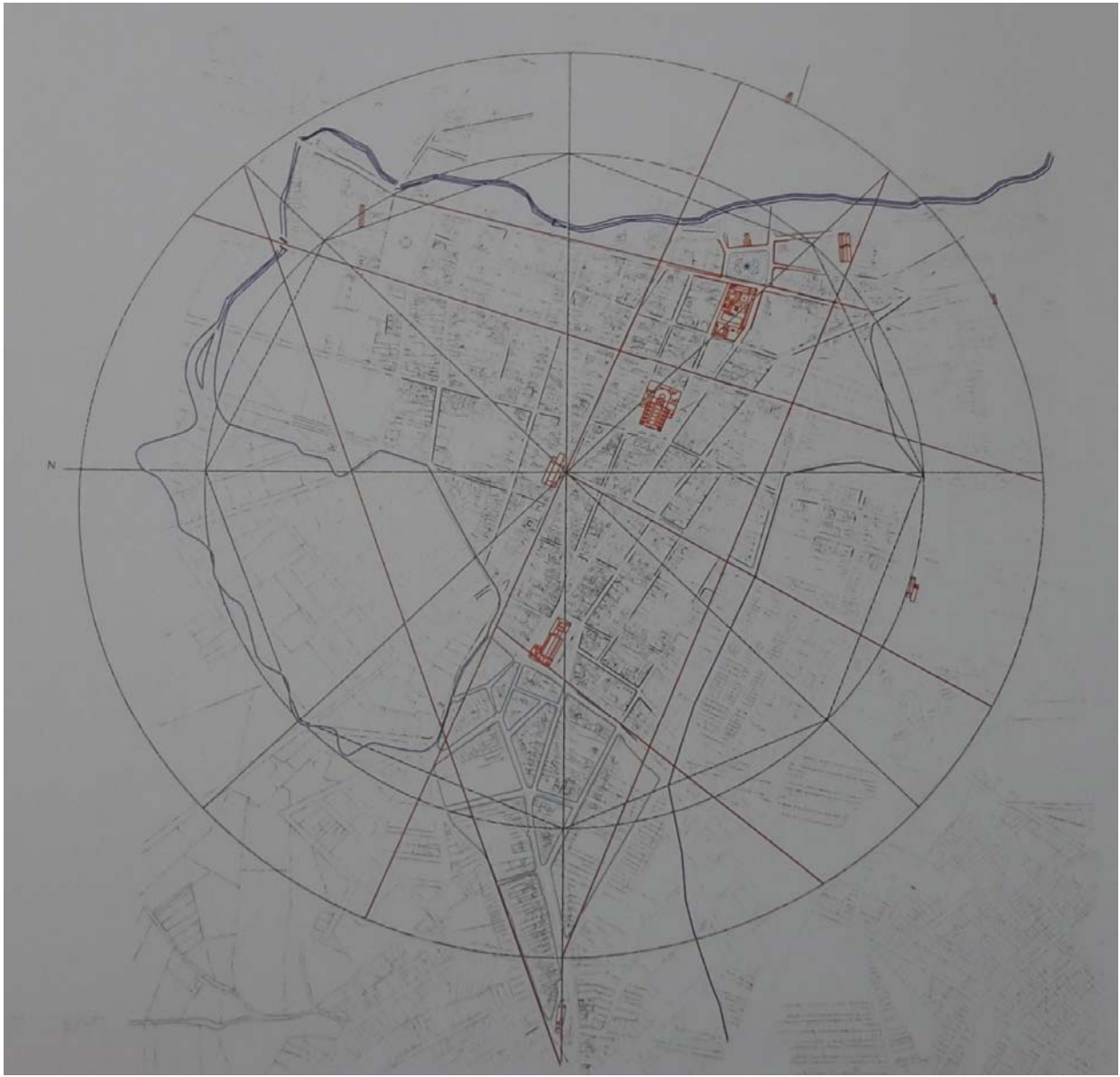

Figure 4: Map drawn by Navarro Segura with the distribution of the streets of San Cristóbal de La Laguna adjusted to a Vitruvian octagon inscribed in the circumference and related to the main winds. Supposedly centred in the church of San Agustín, it shows the location of several religious buildings as symbolic references to the geometric layout. In this map, north is to the left. (Adapted from Navarro Segura 1999, 233.)

However, some years after Navarro Segura's proposal, other ideas concerning the design of the city emerged, which, although they agreed with its geometric origin, differed in some of the details. For example, it was noted that the map of the historic centre did not show clear traces of a design based on the radius of a circumference. This was evidenced by the morphological aspect of the city, as well as by the layout of the streets and the location of the most representative old buildings (Herráiz Sánchez 2007, 61). Furthermore, once the alignment of three prominent constructions (the tower of La Concepción in the old town, the Cathedral, and the chapel of San Miguel) was verified, Herráiz Sánchez hypothesised that there existed a new orientation axis for the city, which would be parallel to the axis of symmetry of La Concepción. This direction indicates roughly the solstitial line. It was also suggested that the relevance of the east-west (equinoctial) axis emphasised by Navarro Segura did not find an adequate justification in a city that, supposedly, was oriented towards a solstitial direction (we will see more details on this point in the next section). Finally, Herráiz Sánchez (2007) suggested that the angle that forms the symmetry axis of La Concepción with respect to cardinal east, that he estimated to be approximately 26°.5, represented the minor angle of a "golden triangle". From this, it was proposed that "the contours of the city, the location of religious buildings and the orientation and dimensions of the main streets are the result of adapting to the orography of the place the geometrical principles of the Golden Section, which coincide in the latitude of La Laguna with the most relevant solar movements" (Herráiz Sánchez 2007, 85).

To conclude this section, let us remember that one of the outstanding subjects for the readers of *De Architectura* was the Vitruvian reflection on the theory of proportions based on the human figure. In the urban planning proposed by Navarro Segura for San Cristóbal de La Laguna, the idea of anthropomorphisation of the city is also present and this sort of human figure has "the head" in the Adelantado's Square, located to the east. The heart of the body would then be in the centre, occupied by the church of Nuestra Señora de los Remedios, later converted into the main parish church and then into the Cathedral. Finally, at the other end of the city, to the west, in the original nucleus of the Upper Town, the feet of the body would be located, in an agricultural area which formed the base and sustenance of the community (Navarro Segura 1999, 82).

In this way, the Adelantado would have applied to La Laguna the project of a city defined in the *Laws*, Plato's last and unfinished work (Navarro Segura 1999, 166). Thus, the new foundation would have been related to an ideal, mythical city, circular like the soul and the universe, which was imagined eighteen centuries before by the great philosopher, characteristic of a utopian human society where disloyalty is banished, which has no external walls, and is formed by concentric circles and endowed with economic, social, and political structures, as had been suggested according to the old text.

**Methods and results**

We will now attempt to falsify the boldest hypotheses of Navarro Segura and Herráiz Sánchez, which, as we have already mentioned, we consider doubtful due to the available historical documentation. We will show that a simpler interpretation of the present design of the city might be possible and that this plan is based on the orientations of its churches. Because of this, and since the city of La Laguna has several characteristics that make it unique (mentioned in the Introduction), in this section, we will develop a complete study of the orientation of the ensemble of its old churches and chapels. This archeoastronomical approach will shed light on the layout of the city as a whole and, perhaps, will allow us to extract new elements that will reveal the main ideas in the origin of its design.

Our present analysis is the continuation of a large-scale project we are carrying out in the Iberian Peninsula and the Canary Islands. In the latter area, we have already focused on the precise orientation of the colonial churches on the island of Lanzarote and, in (Gangui et al. 2014), the reader can find some of our methods of analysis. In the present work, we consider the first systematic study of the religious constructions of the city of San Cristóbal de La Laguna. Our main interest now is to measure their precise orientations, as this study can provide us with an indication of the layout of the city. Moreover, it can also provide us with unique archaeoastronomical data comprising a compact set of old churches whence we can search for pre-European or canonical religious traditions, including astronomical ones, or a mix of both. As with previous works, this could offer us a broader understanding of one key aspect of Canarian culture (Gangui et al. 2016).

We present our data in Table 1 which shows the results of our fieldwork. The identification of the churches is presented along with their geographical location in coordinates and orientation (archaeoastronomical data): the measured azimuth (rounded to ½° approximation) and the angular height of the point of the horizon towards which the altar or the narthex of the church is facing, as well as the derived computed declination corresponding to the central point of the solar disc. The measured height of the horizon was appropriately corrected for atmospheric refraction (Schaefer 1993, 314). When the horizon was blocked, we employed the digital elevation model based on the Shuttle Radar Topographic Mission (SRTM) available at HeyWhatsThat (Kosowsky 2017), which gives angular heights within a ½° approximation.

| NAME (DATE) | L (°/') North | l (°/') West | a (°) (altar) | h (°) (altar) | δ (°) (altar) | a (°) (narthex) | h (°) (narthex) | δ (°) (narthex) | δ (°) (final) | Patron saint date / Orientation |
|---|---|---|---|---|---|---|---|---|---|---|
| Cruz de los Herreros (XIX c.) | 28.4929 | 16.3155 | 3 | 2 | 63½ | | | | 63½ | 3 May / ---- |
| Cruz de Rodríguez Moure (XVIII c.) | 28.4912 | 16.3181 | 17 | 4 | 60½ | | | | 60½ | 3 May / ---- |
| Sto. Domingo de Guzmán (1602) | 28.4860 | 16.3134 | 88½ | 4½ | 3½ | 268½ | 1½ | -1 | -1 | 8 Aug / 18 Mar – 25 Sep |
| San Miguel (1506) | 28.4875 | 16.3131 | 89 | 11½ | 6½ | 269 | 1 | -0½ | -0½ | 29 Sep / 9 Mar – 15 Sep |
| Ntra. Sra. de Gracia (1540) | 28.4751 | 16.3057 | 104 | -1 | -13 | 284 | 4½ | 14 | 14 | 1st Sunday Aug / 18 Apr–5 Aug |
| San Roque (XVIII c.) | 28.4869 | 16.3102 | 104 | -0½ | -13 | 284 | 2 | 13 | 13 | 16 Aug / 24 Apr – 18 Aug |
| Ntra. Sra. de los Remedios (1515) | 28.4891 | 16.3166 | 109 | 3½ | -15 | 289 | 2½ | 17½ | 17½ | 8 Sep / 1 May – 25 Jul |
| Ntra. Sra. de los Dolores (XVIII c.) | 28.4901 | 16.3168 | 110 | 4½ | -15 | 290 | 6 | 20 | 20 | 15 Sep / 20 May – 23 Jul |
| Convento Sta. Clara de Asís y San Juan Bautista (1577) | 28.4896 | 16.3139 | 113 | 10½ | -14½ | 293 | 4 | 22 | 22 | 11 Aug / 21 May – 2 Jul |
| San Agustín (1506) | 28.4906 | 16.3176 | 114½ | 3 | -19½ | 294½ | 4½ | 23½ | 23½ | 28 Aug / 12 Jun |
| Ntra. Sra. de la Concepción (1515) | 28.4903 | 16.3205 | 115½ | 1 | -22 | 295½ | 4 | 24 | 24 | 8 Dec / 12 Jun |
| Conv Sta. Catalina de Siena (1606), dedic. Virgen del Rosario | 28.4877 | 16.3138 | 190 | -0½ | -61 | 10 | 3½ | 63 | 63 | 7 Oct / ---- |
| San Cristóbal (1552) | 28.4838 | 16.3139 | 251 | 3½ | -15 | 71 | 0½ | 16½ | 16½ | 25 Jul / 26 Apr – 28 Jul |
| El Cristo (San Francisco) (1580) | 28.4934 | 16.3124 | 273 | 2½ | 3½ | 93 | 8½ | 1½ | 1½ | 14 Sep / 14 Mar – 9 Sep |
| San Benito Abad (1554) | 28.4902 | 16.3261 | 278½ | 3 | 9 | | | | 9 | 11 Jul / 3 Apr – 21 Aug |
| San Juan Bautista (1582) | 28.4851 | 16.3190 | 286 | 2½ | 15 | | | | 15 | 24 Jun / 21 Apr – 2 Aug |
| Cruz de los Plateros (XVIII c.) | 28.4872 | 16.3177 | 288 | 2 | 16½ | | | | 16½ | 3 May / 6 May – 7 Aug |
| Santísima Trinidad (casa Peraza y Ayala) (1769) | 28.4872 | 16.3159 | 291 | 2½ | 19½ | | | | 19½ | Mobile / 17 May – 25 Jul |
| Cruz de San Francisco o Cruz de los Álamos (XIX c.) | 28.4928 | 16.3141 | 292 | 4½ | 21 | | | | 21 | 3 May / 25 May – 18 Jul |
| San Diego del Monte (1672) | 28.5008 | 16.3279 | 316 | 12 | 45½ | | | | 45½ | 13 Nov / ---- |
| Cruz Verde (1761) | 28.4857 | 16.3146 | 338 | 3½ | 57 | | | | 57 | 3 May / ---- |

Table 1: Orientations for the churches and chapels of the city of La Laguna. For each of them we show the identification (name and most likely date of construction of the building), the geographical latitude and longitude (L and l), the astronomical azimuth (a) taken along the axis of the building towards the altar or − for some of them − towards the narthex (rounded to ½° approximation), the angular height of the horizon (h) in that direction (including the correction due to atmospheric refraction, and also rounded to ½° approximation) and the corresponding resultant declination (δ). Angular heights of the horizon for directions towards the narthex of the buildings were obtained from numerical terrain models. The final column for the declination shows a combination of both δ (altar) and δ (narthex) in order to emphasise its absolute value, as explained in the text. Finally, the Orientation column is computed by estimating the dates (taking into account the year of the construction of each church) when the final declination of the Sun is the one indicated.

We obtained our measurements using a tandem instrument Suunto 360PC/360R which incorporates a clinometer and a compass with a precision of ½° and also by analysing the surroundings (landscape) of each of the buildings. We then corrected the azimuth data according to the local magnetic declination (Natural Resources Canada 2015), getting values always close to 5°08' W for different sites of the city. Our data is the result of several on-site measurements with a single instrument, taking the axes of the churches, from the back of the buildings towards the altars, as our main guide. In some cases, although not in all, as many churches are surrounded by modern buildings, we could verify that the lateral walls were parallel to their axes. By performing a simple error propagation, we estimate the error of our measurements to be around ±3/4° for the resulting declination. However, given the nature of the measurements, some of them done in the middle of the town and surrounded by asphalt, metal, and wires, we prefer to be more conservative and take our estimated error to be ±1° (upper bound). We consider our data is suitable for a statistical study of the monuments' orientations.

In Figure 5, we show the orientation diagram for the churches and chapels. The diagonal lines on the graph indicate, in the eastern quadrant, the extreme values of the corresponding azimuth for the Sun (azimuths of 62°.7 and 116°.6 − continuous lines – which are equivalent to the northern hemisphere summer and winter solstices, respectively) and for the Moon (azimuths of 56°.6 and 123°.6 − dotted lines −which are equivalent to the position of the major lunistices or lunar standstills).

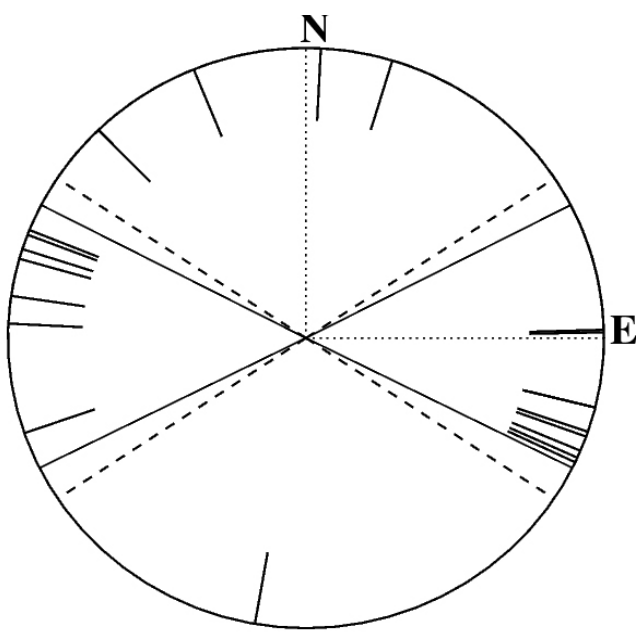

Figure 5: Orientation diagram of the churches and chapels of San Cristóbal de La Laguna, obtained from the data of azimuths in Table 1, considering the direction towards the altar. The majority of the monuments follows the canonical orientation pattern in the solar range, while a few small private chapels concentrate in the northern quadrant and the one of the convent of *Santa Catalina de Siena* points within ten degrees from the south (although its door opens up in the eastward direction).

If we consider the direction, going from the front towards the apse of the churches, of the 21 buildings we measured, seven are oriented in the western quadrant, while nine point towards the eastern quadrant (all of them within the solar range). There are also four which are oriented in the northern quadrant (between 315° and 45°) and only one in the southern quadrant, as shown in Figure 5. Apart from the few small private chapels that are roughly oriented to the north, in general, the vast majority of the constructions are oriented within the solar range.

While there might be different underlying causal factors for this orientation pattern, the idea that the churches may be astronomically-oriented is suggestive. Regarding the solar range, there are a few peculiarities. On the one hand, two of the oldest churches (Santo Domingo and San Miguel) are, to a high degree, oriented towards the equinox. On the other hand, we find two important buildings (apparently) oriented close to the northern hemisphere winter solstice rising Sun, namely the churches of San Agustín and of Nuestra Señora de La Concepción. However, in these cases, the most probable explanation is not given by their orientation towards the eastern horizon, as the winter solstice rising Sun was a very rare target in the Iberian Christian world (González-García and Belmonte 2015), but towards the setting Sun during the opposite (summer) solstice. This would be a reasonable (namely, political and social) solution for the Church to indulge the original population that inhabited those lands (the Guanches) and the first colonists coming to Tenerife from the nearby Gran Canaria island, for whom the summer solstice, contemporary to the time of harvest, was a much more relevant temporal milestone (Belmonte 2015, 1121).

To better understand what we have discussed so far, in Figure 6, we present a declination histogram, corresponding to measurements taken towards the altars, which is independent of geographical location and local topography. This figure shows the astronomical declination versus the normalised relative frequency (to be defined below), which enables a clear and more accurate determination of the structure of the peaks.

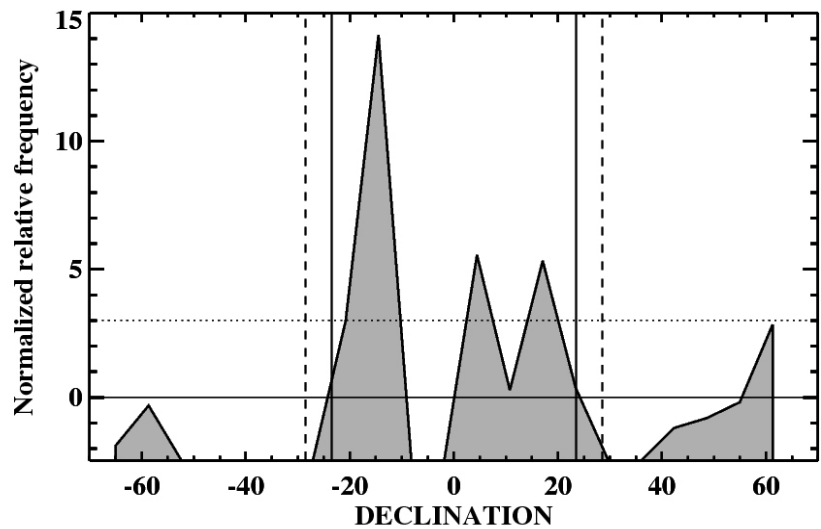

Figure 6: Declination histogram (or curvigram), corresponding to the measurements taken along the axes of the churches towards the altars, showing the estimated "normalised relative frequency", as detailed in the main text. There are three peaks above the 3σ level (dotted horizontal line), with the middle one related to the near-equinoctial direction of some of the buildings. The other two main peaks, although different in height, are symmetrical and are worth further exploration (see the main text). The continuous vertical lines represent the declinations corresponding to the extreme positions of the Sun at the solstices, while the dashed vertical lines represent the same for the Moon in major lunar standstills. In addition to the main peaks, we see two other minor peaks. Both are found close to declination ±60° and might be associated with accumulation peaks due to orientations near the meridian line.

In our analysis, we are using an appropriate smoothing of the declination histogram by a function called "kernel" to generate the kernel density estimate (KDE hereafter). For each entry in declination in Table 1, we multiply the value of the number of occurrences by the kernel function with a given passband or bandwidth (González-García and Belmonte 2014, 100). We have employed an Epanechnikov kernel with a width of twice our estimated error in declination.

Following González-García and Sprajc (2016, 192), to ensure that a concentration of values for the declination is significant, we ought to compare the distribution of our measured data against the result expected from a uniform distribution where the orientations are homogeneously distributed to each possible point in the horizon (the null hypothesis). This comparison will quantify the significance of our results.

Hence, we employ the quantity (f(obs)-f(unif))/σ(unif) for our comparison of declinations, where f(obs) is the frequency of the observed event, f(unif) is the frequency of the uniform event (namely, the frequency of the declinations arising from a sample with uniformly distributed orientations), and σ(unif) is the standard deviation of the uniform distribution. Using this "normalised relative frequency" is equivalent to comparing our data with the results of a uniform distribution of the same size as our data sample and with a mean value equal to the mean of our data (González-García and Sprajc 2016, 195).

Obtaining a KDE-smoothed histogram (a curvigram) scaled with respect to the uniform distribution allows us to see whether our actual data departs significantly from that distribution. As the scale is given by the standard deviation of such uniform distribution, if our data has a maximum that reaches the value 3, for example, it means that it is three times larger than that standard deviation, or 3σ. We take this as the standard criterion to indicate whether our obtained values in declination are significant or not.

In Figure 6, there appear three statistically significant peaks (above the 3σ horizontal line), with the one at the centre, which is located very close to declination zero, probably associated with the equinoctial directions of some of the churches we mentioned before.

However, the other two main peaks of the declination histogram are located at roughly symmetrical positions with respect to the zero-declination equinoctial point, which suggests that these peaks might be related somehow. From the Figure, we clearly see they do not respond to solstitial orientations, so we ought to look for a different explanation. For example, can we be sure the builders of these churches always oriented their constructions in the direction of the rising Sun? What would we find if, due to the orography of the site or for other reasons, they allowed actual orientations to *both* the rising and the setting Sun directions?

In order to check the plausibility of this last hypothesis, we included some measurements of the buildings' axes in the opposite direction to the canonical one, namely towards the narthex of the churches. We show this in the second set of (a, h, δ) data presented in the seventh to ninth columns of Table 1, of which, the "h" data was computed from the digital elevation model based on the SRTM available through HeyWhatsThat (Kosowsky 2017). As we said, although in these kinds of studies, in general, the measurements towards the altar are privileged, the orography of the city and the declination histogram of Figure 6 prompted us test the opposite direction for a few constructions.

In Figure 7, we construct a new declination histogram which allows the possibility of both the rising and setting Sun as orientation targets. The figure shows the astronomical declination (now in absolute value) versus the normalised relative frequency and allows us to see a new structure for the histogram: there appears one outstanding peak dominating the chart which, although it is hard to identify with a single preferred date due to the spread in the data, as reflected in the width of the curve, it might be associated with a date close to the feast-day of San Cristóbal de Licia, the saint to whom La Laguna was originally dedicated.

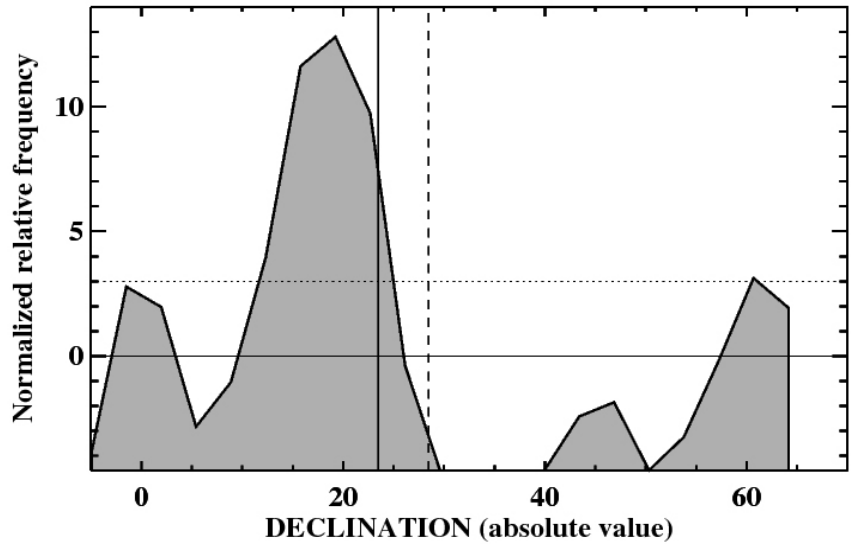

Figure 7: Absolute value declination histogram for the chapels and churches of La Laguna. Only one prominent peak slightly below c. 20° is found above the 3σ level (dotted horizontal line) and may be associated with the Catholic church feast-day of San Cristóbal on July 25$^{th}$ (or 10$^{th}$, both dates being relevant), closely corresponding to the Sun's declination on that day at the time of the founding of the city. As before, the continuous (dashed) vertical line represents the absolute value declination corresponding to the extreme position of the Sun (Moon) at the solstice (at major lunar standstill). In addition to the main peak, we see two barely statistically significant minor peaks. One is found around the equinox, pointing to a canonical orientation pattern, while the other - corresponding to a declination around 60° -, as noted before, might be associated with an accumulation peak due to orientations near the meridian line.

The two declination histograms we have just presented show that those orientations close to the equinoctial direction are not preferred by our measured church data. In the same way, solstitial orientations are also largely absent in these plots. Both these results go against the orientation proposals of previous authors we reviewed in the last section.

As a last note, let us mention that one of the true outstanding features of the city we are studying is its *Latin-cross* structure, roughly depicting a crucified person, which is apparent in the combined layout of some of its most emblematic churches, when one sees the city from above (Figure 8). In fact, a group of five of the oldest churches − still standing today − are so distributed on the land as to suggest the form of a cross, with its head (for example, the inclined head of the crucified Christ) placed in the San Miguel de los Ángeles chapel (just in front of the Adelantado's Square in the Lower Town) and its feet located in the church of La Concepción, in the old part of the city.

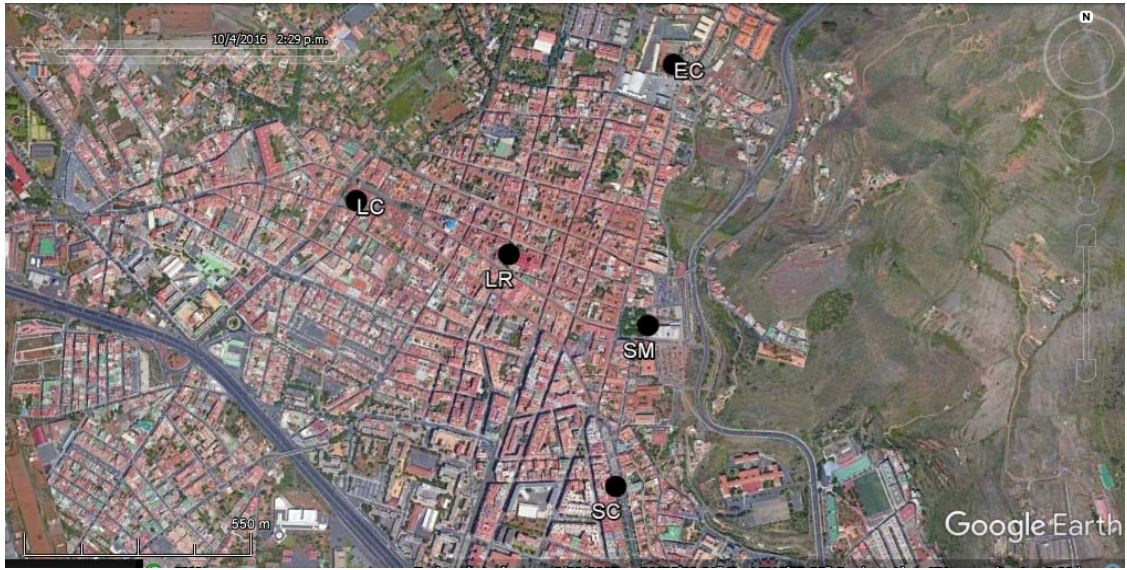

Figure 8: Map of the city of San Cristóbal de La Laguna in Tenerife with the location of five notable religious constructions which display an inclined Latin-cross, generally interpreted as representing a crucified person with the head tilted towards the south. On the map, we can see the church, Nuestra Señora de La Concepción (LC), at the foot of the cross, Nuestra Señora de Los Remedios (LR), the main Cathedral of the city, in the middle, and the San Miguel chapel (SM), located in the Adelantado's Square, at the head of the crucified. On both sides of this head, we find El Cristo (EC, the Saint Francis church, in the north) and the San Cristóbal chapel (SC) in the south. Image based on a map courtesy of Google Earth.

The actual relevance of this singular feature is not clear enough for the time being, as there are no historical sources documenting it. We, therefore, take it as an interesting subject for further research, which we plan to pursue in the near future, given that we do know the characteristics and coordinates of all the relevant buildings. Future work should quantify the statistical significance of the particular alignment of these five emblematic churches, as well as the actual metric of the cross and the involved angles. It is interesting to recall that years before the two recent, and somehow challenging, proposals advanced by the studies of Navarro Segura and of Herráiz Sánchez, which we briefly discussed in the last section, a majority of the inhabitants of La Laguna thought that this obvious *cross* structure was the one underlying the general plan of the city and, hence, it was another reason (or even the *main* reason) for the inclusion of the city as a World Heritage Site by UNESCO.

**Discussion**

In this work we studied the spatial orientation of twenty-one historic Christian churches currently existing in the old part of San Cristóbal de La Laguna to provide new insights on its urban planning. According to the UNESCO declaration, the city exhibits the signs of an interchange of influences between the European and Hispano-Portuguese and American cultures. This feature is apparent not only in its grid plan, but also in its churches and chapels, cloisters, and in the civil architecture, which are closely related to the American ones. La Laguna was the first unfortified Spanish colonial town with a regular plan, and it is thought that its layout provided the model for many colonial towns in the Americas. Designed virtually single-handedly by the Adelantado, Don Alonso Fernández de Lugo, with the grace and support of the Catholic Monarchs, it is outstanding in its planning as a city-territory, supposedly built as a complete and self-contained project as a space for the organization of a new social order (UNESCO 1999). However, as we have also described in this paper, current popular beliefs suggest that the layout of La Laguna would have been designed according to utopian assumptions for an ideal town inspired by Greek philosophical principles.

In our approach to the problem, we took a more pragmatic path towards the study of the actual layout of the city by analysing the exact orientation of its old churches and chapels. We were able to determine that a definite orientation pattern might have been followed and this suggests that the reasons for the actual layout of the city might be simpler than imagined by previous authors. The measured archaeoastronomical data as presented, for example, in the declination histogram of Figure 7, shows a representative pattern which, within certain limits imposed by the limited sample of measured constructions and the resulting width of the main peak, suggests there exists some privileged date which is close to the feast-day of San Cristóbal de Licia, the saint to whom the city was originally dedicated.

We can also see that the prominent, albeit broad, peak of Figure 7 is located around the same declination as that characterising the orientation of the principal church, Nuestra Señora de Los Remedios (declination 17°.5), if one accepts the main alignment of this construction was directed towards the setting Sun. As this church became the main Cathedral of the city, located close to the joining line of the upper and lower villas, its actual orientation could have served as a model to follow for the other churches constructed afterwards. However, in the absence of documentary sources, this hypothesis is difficult to test.

Apart from this, we also found a couple of emblematic churches oriented in the canonical way, namely towards the equinoxes, and another small group aligned with the summer solstice, a characteristic time-mark of the island's aboriginal population before the conquest. However, neither of these two orientations is dominant in the measured sample. This suggests that, given our hypothesis that churches' orientations indicate the original city grid, solstitial or equinoctial directions were not the main target of the builders of the city, contrary to what was proposed in previous investigations of this subject.

To conclude, and based on the foregoing discussion, we think that the planning of the city resulting from the spatial layout of the studied religious buildings could be better understood from simple, well-tested, principles, as the ones employed in this paper, which are ubiquitous in archaeoastronomical research. These include canonical examples of monument orientations, the by-product of Christianization, and a marked preference for aligning religious buildings according to the old town patron saint date. Our findings, at least, momentarily put aside the need for additional hard-to-verify philosophical hypotheses aiming to explain the actual layout of San Cristóbal de La Laguna.